# Chemical-disorder-caused Medium Range Order in Covalent Glass

Xianqiang Liu[1,2]*, Xianbin Li[3], Xinyang Wang[1], Yongqiang Cheng[4], Xuepeng Wang[3], Xiaodong Han[1], Ze Zhang[1,5] & Shengbai Zhang[2,3]†

1 Institute of Microstructure and Property of Advanced Materials, Beijing University of Technology, Beijing 100124, China

2 Department of Physics, Applied Physics, and Astronomy, Rensselaer Polytechnic Institute, Troy, New York 12180, USA

3 State Key Laboratory on Integrated Optoelectronics, College of Electronic Science and Engineering, Jilin University, Changchun 130012, China

4 Chemical and Engineering Materials Division, Oak Ridge National Laboratory, Oak Ridge, Tennessee 37831, USA

5 State Key Laboratory of Silicon Materials and Department of Materials Sciences and Engineering, Zhejiang University, 38 Zheda Road, Hangzhou 310027, China

Correspondence authors: * xqliu@bjut.edu.cn (X.Q. Liu); † zhangs9@rpi.edu (S.B. Zhang)

**Abstract:**

How atoms in covalent solids rearrange over a medium-range length-scale during amorphization is a long pursued question whose answer could profoundly shape our understanding on amorphous ($a$-) networks. Based on *ab-intio* calculations and reverse Monte Carlo simulations of experiments, we surprisingly find that even though the severe chemical disorder in $a$-GeTe undermined the prevailing medium range order (MRO) picture, it is responsible for the experimentally observed MRO. That this thing could happen depends on a novel atomic packing scheme. And this scheme results in a kind of homopolar bond chain-like polyhedral clusters. Within this scheme, the formation of homopolar bonds can be well explained by an electron-counting model and further validated by quantitative bond energy analysis based. Our study suggests that the underlying physics for chemical disorder in $a$-GeTe is intrinsic and universal to all severely chemically disordered covalent glasses.

Understanding amorphous structure is one of the most challenging and long-lasting problems in modern science. Despite the lack of long-range order (LRO), $a$-materials are not completely disordered. In covalent glasses, for example, short-range order (SRO) is preserved as the so-called coordination polyhedral [1], as a result of the directional stereo-chemical bonding. These polyhedra are often organized in some way by sharing corner-, edge-, and even face [1] to exhibit the MRO [1,2]. And this MRO manifests itself as the prepeak (or first sharp diffraction peak) at low wavevectors in the scattering experiments [2]. Up to now, various attractive models have been proposed to decode the MRO of covalent glasses, which include the microcrystalline model, the molecular clusters model [3], the void correlation model [4] and the (nano-) paracrystallite model [2,5-8]. All these models are based on the assumption that the $a$-materials are essentially chemically ordered.

As an important chalcogenide glass and also a crucial component of the GeSbTe family of PCMs [9-15], GeTe is broadly applied in the memory industry for the great electrical and optical contrasts between the $a$- and $c$- states. These property contrasts are a result of the dramatic changes in the atomic arrangements during the phase transitions [16-18]: a considerable amount of the Ge atoms (roughly 1/3) switch from the octahedral (*octa*-) configuration in the crystalline phase to the tetrahedral (*tetra*-) configuration in the amorphous phase [17,18]. Besides, a large amount of Ge-Ge homopolar bonds are observed in *a*-GeTe [19-23]. Even though the homopolar bonds are widely believed to be energetically disfavored, the existence of the Ge-Ge bonds is confirmed by EXAFS experiment [23], which hints for severe chemical disorder in *a*-GeTe.

Different from the prevailing understanding of covalent glass, however, both the chemical order and the coordination polyhedra of *a*-GeTe experience drastic change in comparison with the *crystalline* phase, which not only violates the basic assumption of MRO, but also completely destroys the prevailing picture of MRO in terms of corner-, edge-, face-sharing polyhedral connection. However, the experimental structural factor for GeTe [22] shows clearly the existence of prepeak, which has been interpreted as MRO. This raises the puzzles about the MRO picture in *a*-GeTe. Since MRO is an important topic in amorphous solid that has attracted much attention in recent years for understanding its topology and stability, the answer to the question can be profoundly important to GeTe and to covalent glasses for which the standard MRO definition breakdown.

Here, in *a*-GeTe we propose an atomic packing scheme, termed homopolar bond chain-like polyhedral clusters (HBCPCs), for the chemically disordered amorphous networks to resolve the above controversy. In this scheme, instead of adopting the normal corner- and edge-sharing connections, the Ge-centered tetrahedra interpenetrate each other by sharing Ge-Ge homopolar bonds. This interpenetration is responsible for the observation of MRO. We stress that, in this atomic packing scheme, the Ge-Ge bonds plays a decisive role. Based on an electron-counting model (ECM) [24], we resolve the underlying physics of the chemical disorder in *a*-GeTe. Our bond energy analysis further reveals that the tetrahedral configurations with homopolar bonds are thermodynamically highly competitive with the octahedral configurations without the homopolar bonds.

The AIMD calculations employed the density functional theory with the generalized gradient approximation (GGA) [25], as implemented in the VASP codes [26,27]. The electron-ion interaction was described by the projector augmented wave (PAW) method [28]. It used an energy cutoff of 220 eV for the plane wave expansion, a super-cell of 216 atoms (108 Ge, 108 Te) to mimic the amorphous structure, Γ point for the Brillouin zone sampling, the canonical NVT ensemble with a time step of 3-fs, and the Nose-Hoover thermostat to control the temperature [29,30]. GeTe thin films were directly deposited onto transmission electron microscope (TEM) specimen supporting grids coated with ultrathin carbon films (3 nm) by co-sputtering Ge and Te alloy targets at room temperature. The final thickness of the film was approximately 15 nm. More detailed information can be seen in reference [22]. The electron diffraction (ED) pattern was collected using a JEOL 2010 TEM operating at 200 KV [22]. The experimental S(Q) is obtained from the diffraction intensity profiles. To increase the reliability of the 3D structural model for *a*-GeTe, we also fitted, in the reverse Monte-Carlo (RMC) simulations [31-33], the S(Q) from ED and EXAFS simultaneously (Fig. S1). The simulation box contains 4,000 particles (2,000 Ge plus 2,000 Te), and the density of the GeTe is 0.0337 atom/Å$^3$. The EXAFS backscattering amplitudes and phases were calculated using the FEFF8 code [34] within the self-consistent field approximation.

To develop a comprehensive three-dimensional (3D) model for *a*-GeTe, *ab initio* molecular dynamics (AIMD) calculations were performed. To mimic the severely disordered sputtering system [22], the amorphous structure was obtained by cooling a liquid at 3000K rapidly to 300K and then annealed for 6 pico-second (ps). After the annealing, all the atoms were relaxed to their local minimum-energy positions. The resulting structural model is further validated by comparing with that derived from the *experiment-based* RMC simulations (Fig. S1). Figure 1(a) shows the temperature and free energy evolutions in the AIMD simulation. The top inset in Fig. 1(a) is the liquid (*l*-) state whereas the lower one is the amorphous state. Figure 1(b) shows the typical structure factor S(Q) of the liquid (blue curve, 3000K) and amorphous (red curve, 300K) GeTe states in the AIMD, respectively. At the position indicated by the broken red line (Fig. 1(b)), a prepeak appears in the S(Q) of *a*-GeTe, while it is absent in the liquid state. For a comparison, the experimental S(Q) (black curve), based on the electron diffraction (the top-right inset) of our as-deposit GeTe film, is shown in Fig. 1(b), which agrees well with the AIMD results for amorphous cell, in particular, both curves show a clear prepeak, which suggests that the polyhedra in *a*-GeTe must have maintained some sort of MRO.

In Fig. 1(c), we monitored the time evolution of the bond numbers, in which we used a cutoff = 3 Å. According to Xu *et al*. [35], bond lengths below this cutoff usually have physical meaning while those above it are fake ones. Figure 1(c) shows that after an initial period of little changes, the number of Ge-Te bonds increase rapidly, while those of Ge-Ge and Te-Te decrease accordingly (see the shaded area). After 40 *ps*, the Te-Te bonds are rarely observed. In contrast, a large number of the Ge-Ge bonds remain. It implies that the Ge-Ge bonds are unavoidable and are probably necessary for the structure of *a*-GeTe.

To see the effect of homopolar Ge-Ge bonds on the structure, we examine in Fig. 2 the coordination number distributions (CNDs) and bond angle distributions (BADs) in the *a*-state. Figure 2(a) shows the results for all Ge atoms, where the 3-fold (42.7%) and 4-fold (41.5%) coordinated Ge atoms add up to 84.2% and hence dominate the amorphous network. Figure 2(b) shows the results for Ge atoms only with normal bonds, in which about two thirds of the Ge atoms are 3-fold coordinated with the main bond angle at about 90°. In contrast, Fig. 2(c) shows the results for Ge atoms with at least one homopolar bond. They are dominated by 4-fold coordinated Ge atoms with the main bond angles at about 105°. These results show that Ge atoms only with normal bonds prefer the octahedral configuration, while Ge atoms with homopolar bonds prefer the tetrahedral configuration.

Although the above CND and BAD studies reveal a strong correlation between the tetrahedral configuration and homopolar bonds, the underlying physics is unclear. Here, we will perform an analysis based on the ECM, which shows the coexistence of the *octa*- and *tetra*-Ge in *a*-GeTe, as a manifestation between the multi-valency of Ge atom and the Mott's (8 – N) rule. The original EMC was developed to predict low-energy surface reconstruction of covalent semiconductors [36]. Recently, it was generalized (*g*-ECM) to include the multi-valency to understand edge reconstruction of two-dimensional transitional metal dichalcogenides [24]. Te is a 2 electron acceptor. In the octahedral configuration, CN = 3 for both Ge and Te. To satisfy the charge balance, Ge has to be a 2 electron donor, in spite that it is usually a 4 electron donor. In other words, the two s electrons of Ge do not precipitate in the chemical bonding, which has been explained in terms of the valence alternation model [35]. In the tetrahedral configuration, on the other hand, a Te atom bridges between two adjacent Ge atoms, so the Te coordination is 2 ($CN_{Te}$ = 2). Here, $CN_{Ge}$ = 4 and, by our counting, each Ge is a 4 electron donor. To satisfy the (8-N) rule, therefore each Ge atom either has to bond to 4 bridge-site Te atoms like in a $SiO_2$ or forms two Ge-Te and two Ge-Ge bonds, respectively. The former is impossible because it breaks the 1:1 stoichiometry. The latter is possible as it is allowed by the stoichiometry. Thus, the existence of Ge-Ge bonds in Fig. 1c is necessitated by the tetrahedral configuration. Figure 2(d) shows the two typical configurations of Ge as a result of the AIMD simulations. It reveals that most of the Ge tetrahedra have two Ge-Ge bonds, implying a significant chemical disorder in *a*-GeTe.

From to the above results, we classify the polyhedra in *a*-GeTe into two kinds: the first one is the polyhedral clusters only with normal GeTe bonds. Figure 3(a) shows a snap shot of the network of normal bond polyhedral clusters (NBPCs), in which we have deleted all the Ge-centered polyhedra that have Ge-Ge homopolar bonds. The second one is the HBCPCs. Figure 3(b) shows the same snap shot of HBCPC network. But as the complementary network of NBPCs, only those Ge-centered polyhedra with at least one Ge-Ge bond are kept in the HBCPC network. Further analysis shows that most polyhedra in Fig. 3(a) are defective octahedra. In contrast, most polyhedra in Fig. 3(b) are distorted tetrahedra. Figure 3(c) shows how three NBPCs are connected to form a defective cube by sharing edges and corners of octahedra in the AIMD results in Fig. 3(a), in which the blue circle with green filling is a vacancy. Figure 3(d) shows part of the HBCPC network in Fig. 3(b) in which only two Ge atoms are 3-fold coordinated (indicated by black arrows) while all other Ge atoms are all tetrahedrally coordinated. Importantly, the tetrahedra in Fig. 3(d) tend to interpenetrate each other by sharing Ge-Ge bonds (as indicated by the shaded tetrahedra), in the sense that the central Ge of one tetrahedron is also a corner Ge of an adjacent one. This interpenetration defines the unique atomic packing scheme presented in this work, where the Ge-Ge homopolar bonds serve as the backbones to form the HBCPC network.

The sharing of Ge-Ge bonds between interpenetrated polyhedra allows for the HBCPCs more tightly packed than the NBPCs and, more importantly, it offers an interlocking mechanism between the polyhedra. It can be seen that the octahedra in the NBPC network in Fig. 3(a) are loosely connected, but the ones in the HBCPC network in Fig. 3(b) are more rigidly connected. This qualitative difference between NBPC and HBCPC is expected to result in an imparity in their contributions to MRO.

In order to quantify the difference, we calculate in Fig. 4 the time evolution of S(Q) during an AIMD annealing for the entire $a$-GeTe (middle panel), for NBPCs (left panel), and for HBCPCs (right panel), respectively. The results for $a$-GeTe suggest that the polyhedra start to organize in some orderly way, as the prepeak appears when the temperature is decreased to below 1000K. The results for the HBCPC network show remarkable resemblance to those for entire $a$-GeTe. In contrast, no clear trend of prepeak for the NBPC network: one can say, by inferring to the left panel, that it exists even in the liquid phase, but then disappears and reappears in a totally random fashion during the annealing. Hence, the MRO of $a$-GeTe may be exclusively attributed to the formation of the HBCPCs. This notion is further reinforced by our RMC simulation (see the top part of Fig. 4) where the NBPCs results show no sign of prepeak, but both the HBCPCs and the entire $a$-GeTe results show clear prepeaks.

From the above results, we see that the Ge-Ge homopolar bonds are concomitant with the formation of tetrahedral polyhedra, as a precondition for the formation of the HBCPC network, which is in turn responsible for the MRO observed in $a$-GeTe. The application of $g$-ECM to the amorphous system further shows that the formation of the Ge-Ge bonds in the tetrahedral configuration is physically inevitable. Yet, it can be critically important to know how much it requires to form such a bond from an energetic point of view, because if the energy penalty is too high, none 1:1 ratio should result, although it has not been observed experimentally.

To get a rough idea, we consider the simple NBPC and HBCPC models in Fig. S2 of the supplementary materials. Our first-principles calculation yields the bond energies of $\sigma_{Ge-Ge}^{T-sp^3} = -2.36 eV$ for Ge-Ge, $\sigma_{Ge-Te}^{O-p} = -2.61 eV$ for $p$-bonding Ge-Te, and $\sigma_{Ge-Te}^{T-sp^3} = -2.68 eV$ for $sp^3$-bonding Ge-Te (see Table S1). It is important to note that the Ge-Te bond strength in an octahedral configuration, where $p$-bonding dominates, can be noticeably different from that in a tetrahedral configuration, where $sp^3$-bonding dominates. Thus, the *octa-to-tetra* transition is not simply a replacement of the Ge-Te bond by a Ge-Ge bond that conserves the bond number [37], but a process that also involves the $p$- to $sp^3$-bonding conversion. Taking this effect into account, the energy cost per Ge-Te pair transition is thus

$$\Delta E = (2\sigma_{Ge-Te}^{T-sp^3} + \sigma_{Ge-Ge}^{T-sp^3}) - 3\sigma_{Ge-Te}^{O-p} = 0.11 eV \qquad (1)$$

Such a small energy increase could be easily offset by the $-TS$ term in the Gibbs free energy due to the large entropy ($S$) of the amorphous state.

First-principles calculation, coupled with RMC simulations of experimental data and a $g$-ECM analysis, reveals the formation of Ge-Ge homopolar bonds (which in essence are a chemical disorder) as an intrinsic trend in $a$-GeTe. This is both because of the stoichiometry requirement of the GeTe sample and because of the thermodynamic competitiveness of the tetrahedral configuration with octahedral configuration. It results in the formation of the HBCPC network and the subsequent topological MROs. In chalcogenide glasses, due to the CN mismatch between the cation (usually CN = 4) and anion (usually CN = 2), the higher CN elements, which are typically the cations, must form homopolar bonds among themselves to maintain the stoichiometry. Hence, it should be a general tendency to find cation-centered polyhedra clusters in other severely chemical-disordered covalent glasses, among which chalcogenide glasses are just one example.

Work at RPI was supported by the US Department of Energy (DOE) under Grant No. DE-SC0002623. We also acknowledge the supports by the supercomputer time provided by NERSC under the Grant No. DE-AC02-05CH11231 and the Center of Computational Innovations (CCI) at RPI. XQL acknowledges NSFC (No. 11204008). XBL and XPW acknowledges 973 Program (No. 2014CB921303) and NSFC (No. 11374119).

X.Q. Liu and X.B. Li contribute equally to this work.

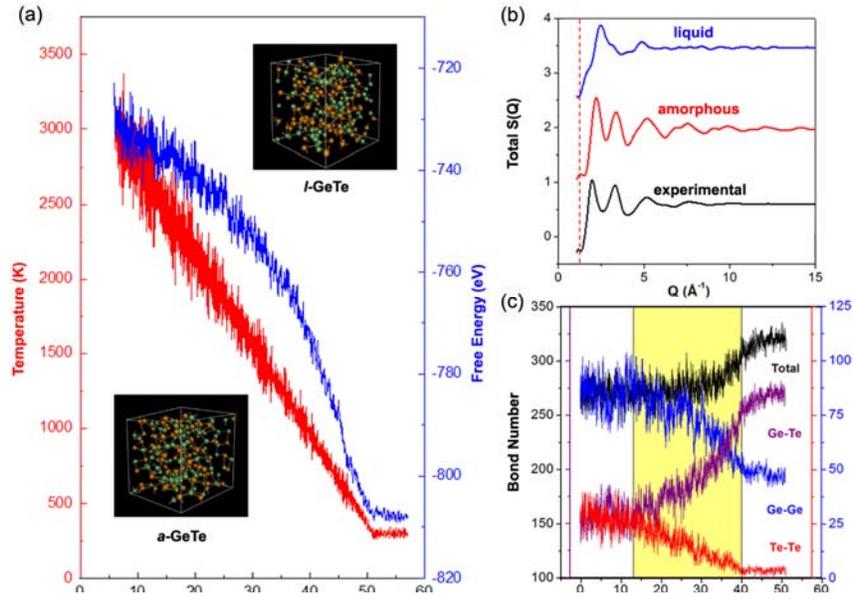

**Figure 1** (a) Time evolution of temperature (red curve) and free energy (blue curve) in AIMD. Upper inset is a snapshot in the liquid (*l*-) state and lower inset is a snapshot in the amorphous (*a*-) state. (b) Structure factors from the *l*- (blue curve, 3000K) and *a*- (red curve, 300K) state AIMD models, respectively. For comparison, the experimental S(Q) (black curve), based on the electron diffraction (the top-right inset) of as-deposit GeTe film, is also shown. (c) Time evolution of bond numbers in the *a*-state.

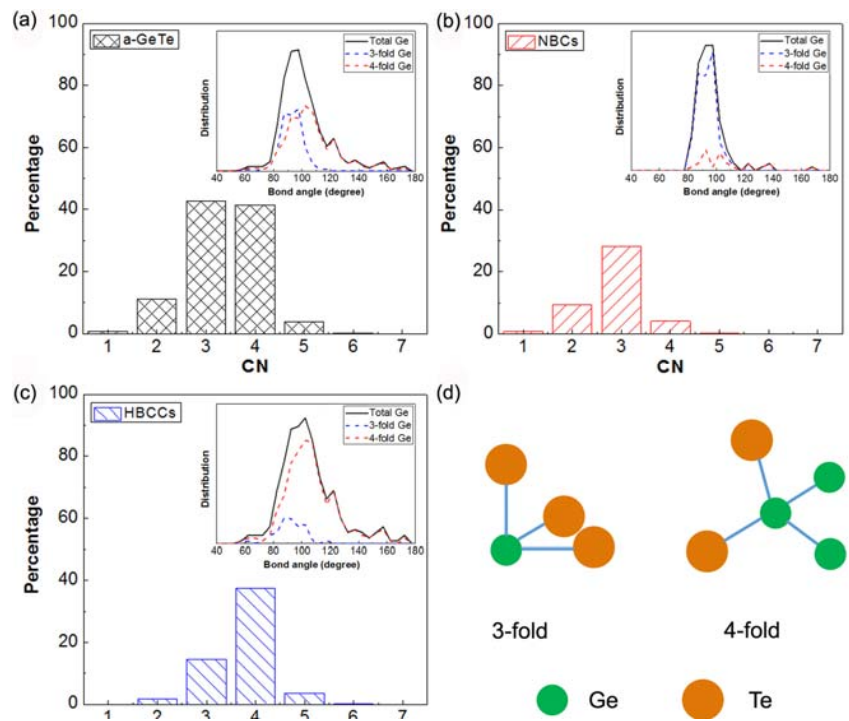

**Figure 2** Coordination number distributions (CNDs) and (insets) bond angle distributions (BADs) of Ge in the *a*-state. (a) For all Ge atoms; (b) for Ge atoms only with normal bonds; (c) for Ge atoms with at least one homopolar bond. (d) Two typical Ge configurations in the *a*-state, obtained by AIMD (orange balls are Te whereas green balls are Ge).

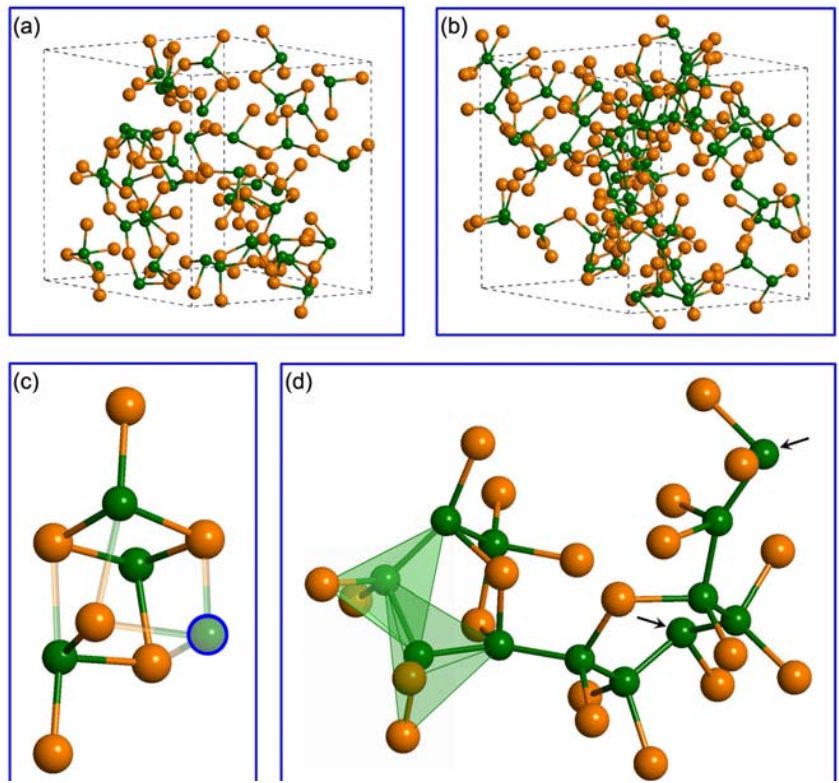

**Figure 3** Atomic structures and networks of NBPCs and HBCPCs in *a*-GeTe. (a) The NBPC network; (b) the HBCPC network; (c) a portion of the NBPC network in Fig. 3(a) where 3 NBPCs and the blue circled "Ge vacancy" site form a "cube"; (d) a portion of the HBCPC network in Fig. 3(b) where polyhedra with homopolar bonds form the (green-color shaded) shared tetrahedra. Arrows indicate 3-fold coordinated Ge atoms.

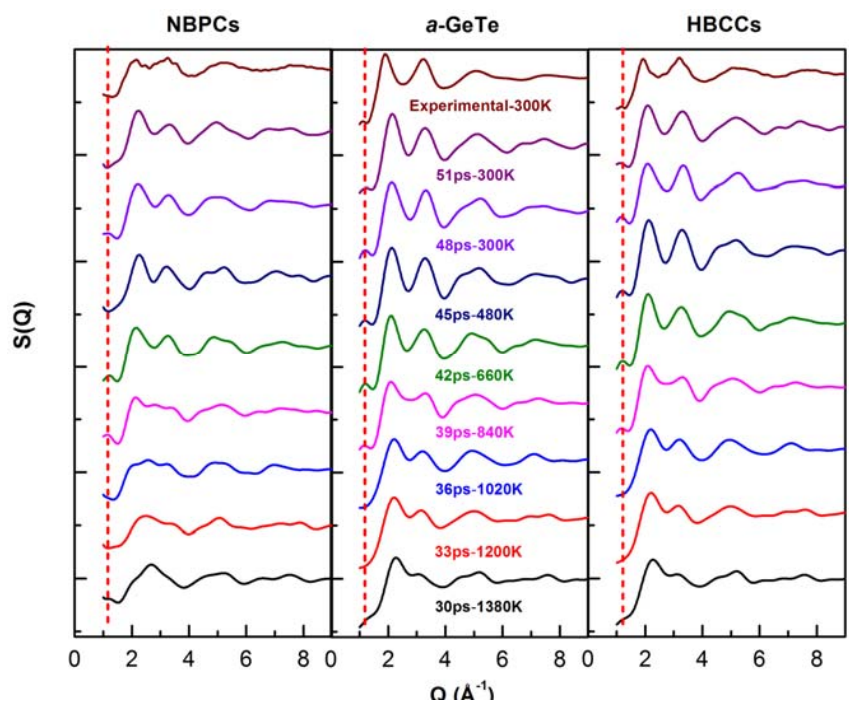

**Figure 4** S(Q) as a function of AIMD simulation time and temperature during annealing. (Middle panel) *a*-GeTe; (left panel) NBPCs; and (right panel) HBCPCs. Starting at 30 *ps*, the data are taken. Subsequent samplings are taken at a 3-*ps* interval. Dashed red line is the prepeak position. For a comparison, S(Q)s from experiment-based RMC modeling are also shown.

16